\documentclass[12pt]{article}
\begin{document}
\begin{center}
\textbf{\large Instanton Cosmology}

\bigskip
Roland E. Allen

Center for Theoretical Physics, Texas A\&M University

College Station, Texas 77843

\bigskip\bigskip\bigskip

\textbf{Abstract}
\end{center}

A new cosmological model implies that  $dR/dt$ is constant for both $t>t_{c}$ 
and $t\rightarrow 0$, and that $dR^{2}/dt \propto n_{s}^{-1}$ for $t_{c}>t>0$. 
Here $R(t)$ is the cosmic scale factor, $t_{c}$ is a crossover time defined in 
the text, and $n_{s}$ is approximately constant near $t_{c}$. The behavior 
at intermediate times is consistent with big-bang nucleosynthesis. The behavior 
at short and long times offers a solution to the smoothness, monopole, 
and flatness problems. In addition, the long-time behavior yields a comfortably 
large age for the universe and predicts that the deceleration parameter 
$q_{0}$ is exactly zero.

\bigskip\bigskip\bigskip\bigskip\bigskip\bigskip
\bigskip\bigskip\bigskip\bigskip\bigskip\bigskip
\bigskip\bigskip\bigskip\bigskip\bigskip\bigskip

\noindent e-mail: allen@tamu.edu \newline 
\noindent fax   : (409) 845-2590 \newline
\noindent tel.  : (409) 845-4341 \newline

\pagebreak

Recently a new fundamental theory was proposed [1]. Among its consequences
is a specific cosmological model, which the present paper will explore in
more detail. We will find that the cosmic scale factor has the following
behavior:

\begin{equation}
R\left( t\right) =\eta \,t\quad ,\quad t\rightarrow 0
\end{equation}
\begin{equation}
R\left( t\right) \propto \left[ \int dt/n_{s}\left( t\right) \right]
^{1/2}\quad ,\quad t<t_{c}
\end{equation}
\begin{equation}
dR\left( t\right) /dt=2\sqrt{3} \quad ,\quad t>t_{c}
\end{equation}
where $\eta \gg 1$ and $n_{s}\left( t\right) $ is a decreasing
function of $t$, with
\begin{equation}
n_{s}(t) \approx \overline{n}_{s} = constant \quad ,\quad \overline{t}<t<t_{c}\;.
\end{equation}
The first prediction implies that the particle horizon in the early universe
is nearly infinite. The second is consistent with the conclusion 
that [2]
\begin{equation}
0.55<\alpha<0.56
\end{equation}
if $R(t) \propto t^{\alpha}$ is assumed to hold during the period of big-bang 
nucleosynthesis [3]. The third implies
that 
\begin{equation}
q=0
\end{equation}
where $q$ is the deceleration parameter in the current epoch.

\bigskip
The theory of Ref. 1 begins with an Ising-like action $S$, which becomes a
Ginzburg-Landau-like action in the continuum approximation. After a Wick
rotation $x^{0}\rightarrow ix^{0}$, one obtains the Lorentzian action 
\begin{eqnarray}
S_{L} &=&iS \\
&=&-\int d^{D}x\left[ \frac{1}{2m}\partial ^{M}\Psi ^{\dagger }\partial
_{M}\Psi -\mu \Psi ^{\dagger }\Psi +\frac{1}{2}b\left( \Psi ^{\dagger }\Psi
\right) ^{2}\right] .
\end{eqnarray}
Here $\Psi $ is a superfield, with both bosonic and fermionic parts. In the
present context, however, it is not necessary to consider fermions or
bosonic excitations. It is also convenient to initially ignore the $d=D-4$
internal dimensions, so that we are left with the action for a GUT Higgs
field in four-dimensional spacetime: 
\begin{equation}
S_{L}~=~-\int d^{4}x~\Psi _{s}^{\dagger }\left( -\frac{1}{2m}\partial ^{\mu
}\partial _{\mu }+\frac{1}{2}bn_{s}-\mu \right) \Psi _{s}
\end{equation}
where $n_{s}=\Psi _{s}^{\dagger }\Psi _{s}$. One can define a 
``superfluid velocity'' 
\begin{equation}
v^{\mu }= -im^{-1}U^{-1}\partial ^{\mu }U
\end{equation}
where 
\begin{equation}
\Psi _{s}=n_{s}^{1/2}U\eta _{0}
\end{equation}
and $\eta _{0}$ is a constant 2-component vector. The field $v_{\alpha
}^{\mu }$, defined by 
\begin{equation}
v^{\mu }=v_{\alpha }^{\mu }\sigma ^{\alpha }\;,
\end{equation}
is interpreted as the vierbein which determines the geometry of spacetime: 
\begin{equation}
e_{\alpha }^{\mu }=v_{\alpha }^{\mu }\;.
\end{equation}
(Here $\sigma ^{0}$ and $\sigma ^{1},\sigma ^{2},\sigma ^{3}$ are
respectively the identity and Pauli matrices.) The curvature of spacetime is
attributed to Planck-scale instantons. This picture leads to both Einstein
gravity and an SO(10) grand unified theory of fermions and the other 
interactions [1].

\bigskip
The condensate $\Psi _{s}$ can support specific kinds of topological
defects, two of which are relevant in the present context: 
(a) SU(2) instantons, which are the four-dimensional generalization of
vortices in an ordinary superfluid. With $x^{0}$ chosen to be the radial
coordinate, $v_{\alpha }^{\mu }$ has the form [4]
\begin{equation}
v_{\alpha }^{k}=\delta _{\alpha }^{k}a/mx^{0}
\end{equation}
\begin{equation}
v_{\alpha }^{0}=0
\end{equation}
where $a$ is a constant and $k=1,2,3$. From a four-dimensional point of
view, it is as if there is a circulation of condensate around the 
origin. (b) U(1) monopole-like defects, which act as sources of the U(1) current 
\begin{equation}
j_{0}^{\mu }=n_{s}v_{0}^{\mu }\;.
\end{equation}
These defects can relieve the constraint 
\begin{equation}
\partial _{\mu }j_{0}^{\mu }=0
\end{equation}
in the same way that vortices in a superfluid can relieve the constraint
\\$\overrightarrow{\nabla}\times\overrightarrow{v_{s}}=0$. Suppose that such
a defect has spherical symmetry, with $j_{0}^{\mu }$ pointing out from the
origin. The equation of continuity (17) still holds for $r\neq 0$, where $r$
is the radial coordinate, so 
\begin{equation}
\partial(r^{3}j_{0})/\partial r=0
\end{equation}
or 
\begin{equation}
v_{0}=a_{1}^{3}/(r^{3}n_{s})
\end{equation}
with $a_{1}$ a constant. From a four-dimensional point of view, it is as if
there is an emanation of condensate from the origin.

\bigskip
These are the two natural point defects for a 2-component field $\Psi _{s}$.
In the present cosmological model they are superposed, and the singularity
at the origin is interpreted as the big bang [1]. Combining (13), (14),
and (19) (with $r\rightarrow x^{0}$) we obtain 
\begin{equation}
e_{\alpha }^{k}=\delta _{\alpha }^{k}a/mx^{0}
\end{equation}
\begin{equation}
e_{\alpha }^{0}=\delta _{\alpha }^{0}a_{1}^{3}/\left( x^{0}\right)
^{3}n_{s}\;.
\end{equation}
The usual relation 
\begin{equation}
e_{\mu }^{\alpha }e_{\beta }^{\mu }=\delta _{\beta }^{\alpha }
\end{equation}
gives the covariant vierbein 
\begin{equation}
e_{k}^{\alpha }=\delta _{k}^{\alpha }mx^{0}/a
\end{equation}
\begin{equation}
e_{0}^{\alpha }=\delta _{0}^{\alpha }\left( x^{0}\right)
^{3}n_{s}/a_{1}^{3}\;.
\end{equation}
In addition, larger values of $x^{0}$ correspond to larger 3-spheres [1],
so \\$dx^{k}=x^{0}d\overline{x}^{k}$, where $d\overline{x}^{k}$ represents a
distance within the unit 3-sphere. There are thus two contributions to the
expansion of the universe in the present model, with both the metric tensor 
$g_{kl}$ and the separation of comoving points increasing as functions of 
$x^{0}$.

\bigskip
The vierbein of (23) and (24) leads to a metric with the Robertson-Walker
form 
\begin{eqnarray}
ds^{2} &=&\eta _{\alpha \beta }\,e_{\mu }^{\alpha }e_{\nu }^{\beta
}\,dx^{\mu }\,dx^{\nu } \\
&=&-\left[ \frac{\left( x^{0}\right) ^{3}n_{s}}{a_{1}^{3}}\right] ^{2}\left(
dx^{0}\right) ^{2}+\left[ \frac{m\left( x^{0}\right) ^{2}}{a}\right] ^{2}
d\overline{x}^{k}d\overline{x}^{k} \\
&=&-dt^{2}+R\left( t\right) ^{2}\left( \frac{dr^{2}}{1-r^{2}}+r^{2}d\Omega
\right) 
\end{eqnarray}
where 
\begin{equation}
dt=\left( n_{s}/4a_{1}^{3}\right) d\left( x^{0}\right) ^{4}
\end{equation}
\begin{equation}
R\left( t\right) =\frac{2m}{a}\left( a_{1}^{3}\int \frac{dt}{n_{s}}\right)
^{1/2}\quad .
\end{equation}
Also, $\eta _{\alpha \beta }=diag\left( -1,1,1,1\right) $ is the Minkowski
metric tensor, and \\$d\Omega =d\theta ^{2}+\sin ^{2}\theta \,d\phi ^{2}$, with 
$\left( r,\theta ,\phi \right) $ representing the usual coordinates of the
unit 3-sphere.

\bigskip
Notice that this metric is determined by the topology of the cosmological
instanton, rather than the Einstein field equations. As usual, these
equations follow from the variational principle $\delta S_{L}/\delta g^{\mu
\nu }=0$ if there are no constraints, where $S_{L}$ is the full Lorentzian
action [1]. In the present model, however, they are subordinate to the
constraint imposed by topology on a cosmological scale. I.e., the Einstein
field equations are only a very good approximation for \textit{local}
deformations of the geometry of spacetime.

\bigskip
The time dependence of the condensate density $n_{s}$ is determined by the
equation of motion for $\Psi _{s}$, which leads to the generalized Bernoulli
equation [1]
\begin{equation}
-\frac{1}{2m}n_{s}^{-1/2}\partial ^{\mu }\partial _{\mu }n_{s}^{1/2}+
\frac{1}{2}mv_{\alpha }^{\mu }v_{\mu }^{\alpha }+bn_{s}=\mu
\end{equation}
or [4]
\begin{equation}
\frac{1}{2m}n_{s}^{-1/2}\left( x^{0}\right) ^{-3}\frac{d}{dx^{0}}\left[
\left( x^{0}\right) ^{3}\frac{d}{dx^{0}}n_{s}^{1/2}\right] +\frac{1}{2}%
mv_{\alpha }^{k}v_{\alpha }^{k}-\frac{1}{2}mv_{\alpha }^{0}v_{\alpha
}^{0}+bn_{s}=\mu \;.
\end{equation}
Substitution of (20) and (21) gives 
\begin{equation}
\xi ^{2}n_{s}^{-1/2}\left( x^{0}\right) ^{-3}\frac{d}{dx^{0}}\left[ \left(
x^{0}\right) ^{3}\frac{d}{dx^{0}}n_{s}^{1/2}\right] +\frac{3\xi ^{2}a^{2}}
{\left( x^{0}\right) ^{2}}-\frac{\xi ^{2}m^{2}a_{1}^{6}}{\left( x^{0}\right)
^{6}n_{s}^{2}}+\frac{n_{s}}{\overline{n}_{s}}=1\;
\end{equation}
where $\xi =\left( 2m\mu \right) ^{-1/2}$ is the coherence length and 
$\overline{n}_{s}=\mu /b$ is the condensate density in a uniform system. With
the scalings 
\begin{equation}
\rho =x^{0}/\xi \quad ,\quad f=\left( n_{s}/\overline{n}_{s}\right) ^{1/2}
\end{equation}
we obtain an equation for $f$ which is analogous to the standard equation
for the condensate density near a superfluid vortex: 
\begin{equation}
\frac{1}{\rho ^{3}}\frac{d}{d\rho }\left( \rho ^{3}\frac{df}{d\rho }\right) +
\frac{3a^{2}}{\rho ^{2}}f-\frac{\overline{a}^{6}}{\rho ^{6}f^{4}}
f+f^{3}-f=0\;
\end{equation}
where 
\begin{equation}
\overline{a}^{3}=a_{1}^{3}m/\overline{n}_{s}\xi ^{2}\;.
\end{equation}
The asymptotic solutions are 
\begin{equation}
f=1-3a^{2}/2\rho ^{2}\quad ,\quad \rho \rightarrow \infty
\end{equation}
\begin{equation}
f=c/\rho \quad ,\quad \rho \rightarrow 0
\end{equation}
with 
\begin{equation}
c^{6}+c^{4}\left( 3a^{2}-1\right) =\overline{a}^{6}\;.
\end{equation}

\bigskip
The forms (36) and (37) imply that there is a $\rho _{c}$ such that 
\begin{equation}
f\left( \rho _{c}\right) =1\;.
\end{equation}
Suppose that 
\begin{equation}
\rho _{c}^{6}\gg \overline{a}^{6}
\end{equation}
and 
\begin{equation}
\overline{a}^{2}\gg 3a^{2}\;.
\end{equation}
One then expects that the first term of (34) is very small at $\rho _{c}$,
and this expectation is confirmed by numerical calculations [5]. If this
term is neglected, (34) and (39) give 
\begin{equation}
\rho _{c}\approx \overline{a}\left( \overline{a}^{2}/3a^{2}\right) ^{1/4}\;.
\end{equation}

\bigskip
One expects $f$ to be a slowly decreasing function of $\rho$ in the region 
\\$\overline{a}<\rho<\rho _{c}$, and this is again confirmed by the
numerical calculations [5]. Then the condensate density $n_{s}$ is a
slowly decreasing function of the time $t$ in a region 
$\overline{t}<t<t_{c}$, 
where $\overline{t}$ and $t_{c}$ are the times corresponding to 
$\rho =\overline{a}$ and $\rho _{c}$: 
\begin{equation}
\overline{t}\sim \left( \overline{n}_{s}/4a_{1}^{3}\right) \xi^{4}\overline
{a}^{4}\sim m\xi ^{2}\overline{a}
\end{equation}
\begin{equation}
t_{c}\sim \left( \overline{n}_{s}/4a_{1}^{3}\right) \xi ^{4}\rho
_{c}^{4}\sim m\xi ^{2}\left( \overline{a}/a\right) ^{2}\overline{a}\;.
\end{equation}
(See (28), (33), and (35). Also notice that (43) is consistent with (54).)
Since $m\xi ^{2}\sim t_{P}$, where $t_{P}\sim 10^{-42}$ sec is a Planck 
time [1], we can also write 
\begin{equation}
\overline{t}\sim \overline{a}\,t_{P}
\end{equation}
\begin{equation}
t_{c}\sim \overline{\eta }^{2}\overline{t}
\end{equation}
where 
\begin{equation}
\overline{\eta}=\overline{a}/a \gg 1\;.
\end{equation}
For example, $\overline{t}\sim 100$ sec  and 
$\overline{\eta }$ $\sim 100$ corresponds to $t_{c}\sim 10^{6}$ sec.

\bigskip
Starting with the action $S_{L}=\int dx^{0}\,L$ of (9), one can follow the
usual rules for obtaining the Hamiltonian $H=\int d^{3}x\,\,\mathcal{H}$.
The result is 
\begin{equation}
\mathcal{H}=\overline{n}_{s}\mu f\left[ -\frac{1}{\rho ^{3}}\frac{d}{d\rho }
\left( \rho ^{3}\frac{df}{d\rho }\right) +\frac{3a^{2}}{\rho ^{2}}f+\frac
{\overline{a}^{6}}{\rho ^{6}f^{4}}f+\frac{1}{2}f^{3}-f\right] \;.
\end{equation}
Since the volume of a 3-sphere is 
$2\pi^{2} \left( x^{0}\right) ^{3}=$ $2\pi^{2} \xi^{3}\rho ^{3}$, 
(37) and (38) imply that 
\begin{equation}
H=\pi^{2} \overline{n}_{s}\mu \xi ^{3} \left( 3c^{4}+12c^{2}a^{2}\right) /\rho
\approx 3\pi^{2} \overline{n}_{s}\mu \overline{a}^{4}\xi ^{3}/\rho
\end{equation}
as $\rho \rightarrow 0$. (According to (2.5)-(2.7) of Ref. 1, there is a
cutoff at $x^{0}\sim \xi $, or $\rho \sim 1$, so that the initial value of 
$H$ is comparable to $\overline{a}^{4}m_{P}$.) In this same regime, (28),
(29), (33), and (35) give 
\begin{equation}
t=\left( c^{2}m\xi ^{2}/2\overline{a}^{3}\right) \rho ^{2}
\end{equation}
\begin{equation}
n_{s}=\left( c^{4}m\xi ^{2}/2\overline{a}^{3}\right) \overline{n}_{s}/t
\end{equation}
\begin{equation}
R\left( t\right) =\eta t
\end{equation}
where
\begin{equation}
\eta =2\overline{a}^{3}/c^{2}a\approx 2\overline{\eta }\gg 1
\end{equation}
with each equation holding in the limit $\rho \rightarrow 0$, or $
t\rightarrow 0$. It follows that 
\begin{equation}
t\approx \frac{m\xi ^{2}}{2\overline{a}}\rho ^{2}\sim \frac{\left( \rho
t_{P}\right) ^{2}}{\overline{t}}
\end{equation}
\begin{equation}
n_{s}\approx \frac{\overline{a}m\xi ^{2}}{2}\frac{\overline{n}_{s}}{t}\sim 
\frac{\overline{t}}{t}\overline{n}_{s}
\end{equation}
\begin{equation}
R\left( t\right) \approx \eta t
\end{equation}
for $t<\overline{t}$.

\bigskip
The particle horizon is determined as usual by 
\begin{equation}
d_{H}\left( t\right) =R\left( t\right) \int_{{\epsilon }}^{\,{\,t}}\frac{
dt^{\prime }}{R\left( t^{\prime }\right) }\quad ,\quad \epsilon \rightarrow
0\;.
\end{equation}
In Friedmann models with $R\left( t\right) \propto t^{\alpha }$, and $\alpha
<1$, the horizon is finite:

\begin{equation}
d_{H}\left( t\right) =t\,/\left( 1-\alpha \right) \sim t\;.
\end{equation}
In the present model, however, (52) implies that $\alpha \approx 1$ in the
early universe, making the horizon nearly infinite: for $t<\overline{t}$,

\begin{equation}
d_{H}\left( t\right) \approx t\log \left( t/\epsilon \right) \rightarrow
\infty \;.
\end{equation}

\bigskip
Now let us turn to the behavior of the condensate at large times \\$t>t_{c}$.
This is the regime already considered in Ref. 1, where the following
assumptions are made: (1) There are monopole-like defects which relieve the
constraint (17) by acting as sources of the current $j_{0}^{\mu }$, in the
same way that vortices permit a superfluid to rotate by relieving the
constraint \\$\overrightarrow{\nabla }\times \overrightarrow{v_{s}}=0$. (See
the discussion following (7.24) of Ref. 1.) These are Planck-scale versions
of the cosmological U(1) monopole-like defect considered above. (2) At large
times ($t>t_{c}$), the condensate density is constant in external spacetime.
This requires that 
\begin{equation}
\frac{1}{2}mv_{\alpha }^{k}v_{\alpha }^{k}-\frac{1}{2}mv_{\alpha
}^{0}v_{\alpha }^{0}=0\quad ,\quad t>t_{c}\;.
\end{equation}
(See the discussion above (5.14) and below (8.40) in Ref. 1.) It follows
that 
\begin{equation}
v_{\alpha }^{0}v_{\alpha }^{0}=v_{\alpha }^{k}v_{\alpha }^{k}=3\left(
a/mx^{0}\right) ^{2}
\end{equation}
or 
\begin{equation}
e_{0}^{\alpha }=\delta _{0}^{\alpha }mx^{0}/\sqrt{3}a\;.
\end{equation}

\bigskip
Let us now consider the origin of the condition (60) in more detail, by
including the internal space $x_{B}$ of Ref. 1, with $d$ coordinates $x^{m}$%
, as well as our external space with coordinates $x^{\mu }$. The condensate
wavefunction can always be written in the form 
\begin{equation}
\Psi _{s}\left( x_{A},x_{B}\right) =\Psi _{A}\left( x_{A}\right) \Psi
_{B}\left( x_{A},x_{B}\right) 
\end{equation}
with 
\begin{equation}
n_{s}\left( x_{A},x_{B}\right) =n_{A}\left( x_{A}\right) n_{B}\left(
x_{A},x_{B}\right) 
\end{equation}
\begin{equation}
n_{A}=\Psi _{A}^{\dagger }\Psi _{A}\quad ,\quad n_{B}=\Psi _{B}^{\dagger
}\Psi _{B}\;.
\end{equation}
For $t>t_{c}$ it is assumed that the condensate has settled into a local
energy minimum, with an internal instanton satisfying 
\begin{equation}
\left[ -\frac{1}{2m}\partial ^{m}\partial _{m}+V\left( x_{B}\right) -\mu
\right] \Psi _{B}\left( x_{B}\right) =0
\end{equation}
\begin{equation}
V\left( x_{B}\right) =bn_{A}n_{B}\left( x_{B}\right) \;.
\end{equation}
(See (5.14) of Ref. 1, and the comment below (8.40).) At all times, on the
other hand, one has the equation of motion 
\begin{equation}
\left( -\frac{1}{2m}\partial ^{\mu }\partial _{\mu }-\frac{1}{2m}\partial %
^{m}\partial _{m}+bn_{A}n_{B}-\mu \right) \Psi _{A}\Psi _{B}=0\;.
\end{equation}
(See (3.7) of Ref. 1.) In external spacetime this becomes 
\begin{equation}
\left( -\frac{1}{2m}\partial ^{\mu }\partial _{\mu }+\widetilde{b}n_{A}-%
\widetilde{\mu }\right) \Psi _{A}=0
\end{equation}
where 
\begin{equation}
\widetilde{\mu }=\mu -T_{B}
\end{equation}
\begin{equation}
T_{B}=N_{B}^{-1}\int d^{d}x\;\Psi _{B}^{\dagger }\left( -\frac{1}{2m}
\partial ^{m}\partial _{m}\right) \Psi _{B}
\end{equation}
\begin{equation}
\widetilde{b}=bN_{B}^{-1}\int d^{d}x\;n_{B}^{2}
\end{equation}
\begin{equation}
N_{B}=\int d^{d}x\;n_{B}\;.
\end{equation}
It is reasonable to make the approximation that $\widetilde{\mu }$ and 
$\widetilde{b}$ are slowly varying functions of $t$. We then regain the
action of (9) and the previous results for $t<t_{c}$, with the replacements 
$\mu \rightarrow \widetilde{\mu }$, $b\rightarrow \widetilde{b}$, and 
$n_{s}\rightarrow n_{A}$.

\bigskip
In particular, the scaled equation (34) still holds as $t\rightarrow t_{c}$
from below. Since $t_{c}$ is defined to be the point where $f=1$, and the
condensate is assumed to remain fully formed beyond this point, (33) implies
that
\begin{equation}
n_{A}=\overline{n}_{A}=\widetilde{\mu }/\widetilde{b}\quad \mbox{for}\quad t
\geq t_{c}
\end{equation}
with $\widetilde{\mu }$ and $\widetilde{b}$ constant. Then (69) gives
\begin{equation}
\Psi _{A}^{\dagger }\left( -\frac{1}{2m}\partial ^{\mu }\partial _{\mu
}\right) \Psi _{A}=0\;.
\end{equation}
The arguments in Section 3 of Ref. 1 show that this condition is equivalent
to
\begin{equation}
\frac{1}{2}mv_{\alpha }^{\mu }v_{\mu }^{\alpha }=0
\end{equation}
which is the same as (60).

\bigskip
With $e_{0}^{\alpha }$ given by (62), and $e_{k}^{\alpha }$ still given by
(23), (25) becomes
\begin{eqnarray}
ds^{2} &=&-\left( \frac{mx^{0}}{\sqrt{3}a}\right) ^{2}\left( dx^{0}\right)
^{2}+\left[ \frac{m\left( x^{0}\right) ^{2}}{a}\right] ^{2}d\overline{x}^{k}
d\overline{x}^{k} \\
&=&-dt^{2}+R\left( t\right) ^{2}\left( \frac{dr^{2}}{1-r^{2}}+r^{2}d\Omega 
\right) 
\end{eqnarray}
where
\begin{equation}
dt=\left( m/2\sqrt{3}a\right) d\left( x^{0}\right) ^{2}
\end{equation}
\begin{equation}
R\left( t\right) =m\left( x^{0}\right) ^{2}/a\;.
\end{equation}
It follows that
\begin{equation}
\stackrel{\cdot }{R}=dR/dt=2\sqrt{3}\;.
\end{equation}
With the conventional definitions

\begin{equation}
H=\,\stackrel{\cdot }{R}/R
\end{equation}
or
\begin{equation}
q=-\stackrel{..}{R}/RH^{2}
\end{equation}
we have
\begin{equation}
q=0\quad ,\quad t>t_{c}\;.
\end{equation}

\bigskip
The present model can be viewed from either of two perspectives:

\bigskip
In the Lorentzian spacetime of human observers, the universe is born
containing a highly confined and coherent bosonic field $\Psi _{s}$, which
is analogous to the coherent photon field in a laser cavity. The density 
${n}_{s}$ of this field is initially very large (at $x^{0}\sim t_{P}$), but
decreases to its minimum-energy value $\overline{n}_{s}$ during a period 
$t<t_{c}$. At the same time, it releases energy to other fields, producing a
hot big bang (although the coupling to other fields is omitted in the
present treatment). For $t>t_{c}$, $\Psi _{s}$ remains fully condensed, and
can be properly interpreted as the condensate associated with a GUT Higgs
field.

\bigskip
In Euclidean spacetime, which is regarded as more fundamental in Ref. 1, the
big bang results from a topological defect centered on the point $t=0$. The
nature of this cosmological instanton is specified above (20). When one
transforms the solution (37) to Euclidean spacetime, the action $S$ is found
to be positive, as is the Hamiltonian $H$ in Lorentzian spacetime. $S$ and $H
$ are also positive for the Planck-scale monopole-like defects postulated
above (60).

\bigskip
Let us now consider the physical implications of the above results:

\bigskip
The fact that the particle horizon is nearly infinite at early times means
that the observable universe was once causally connected, so its isotropy
and homogeneity are not unnatural, as they are in standard cosmology without
inflation [6-11].

\bigskip
In addition, since the Kibble mechanism requires a horizon [12], there is
no longer a compelling mechanism for the production of unacceptably large
quantities of cosmic relics like monopoles.

\bigskip
The curvature associated with (78) is proportional to $R\left( t\right) ^{-2}
$ [13]. The present model thus requires a universe that is asymptotically
flat, because the topology of the cosmological instanton requires that 
$R\left( t\right) \propto t$ for $t\rightarrow \infty $. This is true even if

\begin{equation}
\Omega _{m}=\rho _{m}/\rho _{m}^{c}\neq 1
\end{equation}
where $\rho _{m}^{c}$ is the critical mass density corresponding to a flat
universe in conventional cosmology. Observations of the cosmic microwave
background and of large scale structure indicate that spacetime is very
nearly flat but that $\Omega _{m}\approx 0.3$ [14].

\bigskip
If the period $t<t_{c}$ is neglected, (81) implies that $R\left( t\right)
\propto t$, so $H^{-1}=t$ rather than $\frac{3}{2}t$. The age of the
universe is then given by $t_{0}=H_{0}^{-1}$ rather than $\frac{2}{3}%
H_{0}^{-1}$. With the conventional definition $H_{0}=100h$ km s$^{-1}$Mpc$%
^{-1}$, a value of $h=0.65$ [15] implies that $t_{0}=15$ Gyr rather than 
$10$ Gyr. This larger value is more consistent with recent estimates of $%
10-14$ Gyr for the ages of the oldest globular clusters [16].

\bigskip
Perhaps the most directly testable prediction of the present theory is 
\\$q=0$,
which means that the universe is neither decelerating nor accelerating [17]. 
This feature clearly demarcates the present theory from standard cosmology 
with a flat or closed universe [18,19], in which $q_{0}\geq 1/2$, and from 
$\Lambda$CDM models [20], in which $q_{0}\sim -1/2$. Quantitative
measurements of $q_{0}$ are now becoming attainable [21,22], so it should
be possible to subject this prediction to convincing tests in the near
future.

\pagebreak

\noindent
\textbf{Acknowledgement} 

\bigskip
This work was supported by the Robert A. Welch Foundation. I have greatly
benefitted from conversations with R.L. Arnowitt on new observational
developments, and would also like to thank G. Steigman  for informing me of
the results of Ref. 2 before it was submitted for publication.

\pagebreak

\noindent
\textbf{References}

\bigskip\noindent
[1]  R.E. Allen, Int. J. Mod. Phys. A \textbf{12} (1997) 2385,
hep-th/9612041.

\bigskip\noindent
[2]  M. Kaplinghat, G. Steigman, I. Tkachev, and T.P. Walker,
astro-ph/9805114.

\bigskip\noindent
[3]  D.N. Schramm, \textit{The Big Bang and Other Explosions in Nuclear
and Particle Astrophysics }(World Scientific, Singapore, 1996).

\bigskip\noindent
[4]  The $v_{k}^{\alpha }$ are ``ordinary'' components, after the initial
coordinates of Ref. 1 are transformed to four-dimensional spherical
coordinates. See S. Weinberg, \textit{Gravitation and Cosmology} (Wiley, New
York, 1972), p. 109.

\bigskip\noindent
[5]  R.E. Allen, J. Dumoit, and A. Mondragon, to be published in the 
proceedings of PASCOS-98 (Sixth International Symposium on Particles, 
Strings and Cosmology; Northeastern University, Boston; March 22-27, 1998).

\bigskip\noindent
[6]  P.J.E. Peebles, \textit{Principles of Physical Cosmology} (Princeton
University Press, Princeton, 1993), and references therein.

\bigskip\noindent
[7]  E.W. Kolb and M.S. Turner, \textit{The Early Universe}
(Addison-Wesley, Redwood City, California, 1990), and references therein.

\bigskip\noindent
[8]  \textit{The Early Universe: Reprints}, edited by E.W. Kolb and M.S.
Turner (Addison-Wesley, Redwood City, California, 1988).

\bigskip\noindent
[9]  P.D.B. Collins, A.D. Martin, and E.J. Squires, \textit{Particle
Physics and Cosmology} (Wiley, New York, 1989).

\bigskip\noindent
[10]  A. Linde, \textit{Particle Physics and Inflationary Cosmology}
(Harwood, Chur, Switzerland, 1990).

\bigskip\noindent
[11]  \textit{Unsolved Problems in Astrophysics,} edited by J.N. Bahcall
and J.P. Ostriker (Princeton University Press, Princeton, 1997).

\bigskip\noindent
[12]  A. Vilenkin and E.P.S. Shellard, \textit{Cosmic Strings and Other
Topological Defects} (Cambridge University Press, Cambridge, 1994).

\bigskip\noindent
[13]  C.W. Misner, K.S. Thorne, and J.A. Wheeler, \textit{Gravitation}
(W.H. Freeman, San Francisco, 1973).

\bigskip\noindent
[14]  M. Webster, M.P. Hobson, A.N. Lasenby, O. Lahav, and G. Rocha,
astro-ph/9802109, and references therein.

\bigskip\noindent
[15]  W. L. Freedman, J. R. Mould, R. C. Kennicutt, and B. F. Madore, 
astro-ph/9801080, and references therein.

\bigskip\noindent
[16]  See, for example, B. Chaboyer, P. Demarque, P. J. Kernan, and L. M.
Krauss, astro-ph/9706128.

\bigskip\noindent
[17] Other models predict $q_{0}=0$ for very different physical reasons. 
The first extensive study of such a model was by E.W. Kolb, 
Astrophys. J. \textbf{344} (1989) 543. See also M. Kamionkowski 
and N. Toumbas, Phys. Rev. Lett. \textbf{77} (1996) 587, and references 
therein.

\bigskip\noindent
[18]  Ya.B. Zel'dovich and I.D. Novikov, \textit{Relativistic Astrophysics}
(University of Chicago Press, Chicago, 1983).

\bigskip\noindent
[19]  A.R. Sandage, R.G. Kron, and M.S. Longair, \textit{The Deep Universe} 
(Springer, Berlin, 1995).

\bigskip\noindent
[20]  M.S. Turner, Nucl. Phys. \textbf{A621} (1997) 522c, astro-ph/9704024, 
and references therein. 

\bigskip\noindent
[21] S. Perlmutter, G. Aldering, M. D. Valle, S. Deustua, R.S. Ellis, S.
Fabbro, A. Fruchter, G. Goldhaber, A. Goobar, D.E. Groom, I. M. Hook, A.G.
Kim, M.Y. Kim, R.A. Knop, C. Lidman, R. G. McMahon, P. Nugent, R. Pain, N.
Panagia, C.R. Pennypacker, P. Ruiz-Lapuente, B. Schaefer, and N. Walton, 
Nature \textbf{391} (1998) 51, astro-ph/9712212.

\bigskip\noindent
[22]  P. M. Garnavich, R. P. Kirshner, P. Challis, J. Tonry, R. L.
Gilliland, R. C. Smith, A. Clocchiatti, A. Diercks, A. V. Filippenko, M.
Hamuy, C. J. Hogan, B. Leibundgut, M.M. Phillips, D. Reiss, A. G. Riess, B.
P. Schmidt, J. Spyromilio, C. Stubbs, N. B. Suntzeff, and L. Wells, 
Astrophys. J. \textbf{493} (1998) L53, astro-ph/9710123.

\end{document}